\documentclass[referee]{raa}            

\usepackage{graphicx,times}             

\begin{document}

   \title{Fine-scale structures and material flows of quiescent filaments observed by New Vacuum Solar Telescope\,$^*$
   \footnotetext{$*$ Supported by the National Natural Science Foundation of China.}}

   \volnopage{Vol.0 (200x) No.0, 000--000}      
   \setcounter{page}{1}          

   \author{Xiao-Li Yan\inst{1,2}
   \and Zhi-Ke Xue\inst{1}
   \and Yong-Yuan Xiang\inst{1}
   \and Li-Heng Yang\inst{1}}

   \institute{Yunnan Observatories, Chinese Academy of Sciences, Kunming 650011, China; {\it yanxl@ynao.ac.cn}\\
        \and
            Key Laboratory of Solar Activity, National Astronomical Observatories, Chinese Academy of Sciences, Beijing 100012, China.}

   \date{Received~~2013 month day; accepted~~2013~~month day}

\abstract{Study on the small-scale structures and material flows of solar quiescent filaments is very important for understanding the formation and equilibrium of solar filaments. Using the high resolution H$\alpha$ data observed by the New Vacuum Solar Telescope (NVST), we present the structures of the barbs and the material flows along the threads across the spine in two quiescent filaments on 2013 September 29 and on 2012 November 2, respectively. During the evolution of the filament barb, several parallel tube-shaped structures formed and the width of the structures ranges from about 2.3 Mm to 3.3 Mm. The parallel tube-shaped structures merged together accompanied with the material flows from the spine to the barb. Moreover, the boundary between the barb and surrounding atmosphere is very neat. The counter-streaming flows were not found to appear alternately in the adjacent threads of the filament. However, the large-scale patchy counter-streaming flows are detected in the filament. The flows in one patch of the filament have the same direction and the flows in the adjacent patch have opposite directions. The patches of two opposite flows with a size of about ten arcseconds exhibited alternately along the spine of the filament. The velocity of these material flows ranges from 5.6 km/s to 15.0 km/s. The material flows along the threads of the filament did not change their direction for about two hours and fourteen minutes during the evolution of the filament. Our results confirm that the large-scale counter-streaming flows with the certain width along the threads of the solar filaments exist and are well coaligned with the threads.
}

   \authorrunning{X. L. Yan et al.}            
   \titlerunning{Fine-scale structures and material flows of quiescent filaments }  

   \maketitle

%
%
\section{Introduction}           

Solar prominences/filaments are relatively cool and high density structures, which are embedded in the chromosphere and the million-degree corona (Hirayama 1985; Tandberg-Hanssen 1995). Prominences appear as the bright structures against the dark background above the solar limb. When projected on the solar disk, they exhibit darker structures than their surroundings in chromospheric lines.

Filaments are typically long-lived features, ranging from about one day to several weeks. Filaments exhibit many different characteristics from sunspots in a solar cycle (Li et al. 2010; Kong et al. 2014; Kong et al. 2015). In general, quiescent filaments have two linked categories of structures: ``spines" and ``barbs". A ``spine" is the highest, horizontal axis part of the quiescent prominence, and is composed of many resolvable threads, while ``barbs" connect to the spine and terminate in the chromosphere (Martin 1998). Moreover, the filament threads in quiescent filaments are shorter than those in active region filaments (Zhou et al. 2014).

The counter-streaming flows along closely spaced vertical regions of a prominence between its top and the lower solar atmosphere is reported by Zirker et al. (1998). It was well-established observationally that the entire filament bodies are composed of numerous thin threads, which are largely horizontal and at angles of 20-25 degrees relative to the long axis of the filament (Engvold et al. 2001). The filament threads may represent separate flux tubes (Engvold 1998). Lin et al. (2003) found that the net counter-streaming flow velocities in the two directions are about 8 km/s. The width of the thin threads is $\leq$ 0.3 arcsec and the velocity of the continuous flow along the threads is about 15 km/s (Lin et al. 2005). The small-amplitude wave is propagating along a number of filament threads with an average phase velocity of 12 km/s and a wavelength of 4 aresecs and the oscillatory period of individual threads varies from 3 to 9 minutes (Lin et al. 2007). By using the off-band H$\alpha$ observation of Hida/DST, Schmieder et al. (2008) reported that the counter-streaming flows in the filament were detected before its eruption and its velocity is 10 km/s. The merger, separation, and oscillation of the filament spines were found by Ning et al. (2009a). Cao et al. (2010) found that the upflows at the filament footpoint were driven by the oscillation of solar surface. The dip model of the prominences claims that the barb is supported by means of a dip in the magnetic field lines over the minor polarity (Aulanier \& Demoulin 1998; van Ballegooijen 2004). Chae et al. (2005) found that the terminating points of the barbs occurred above the minor polarity inversion line by comparing the H$\alpha$ images with the magnetograms taken by SOHO/MDI and the flux cancellation proceeded on the polarity inversion line. Their results are in accordance with the idea that filament barbs are cool plasma suspended in the local dips of the magnetic field lines formed by magnetic reconnection in the chromosphere. The formation and the disappearance of filament barbs may be connected with the flows (Joshi et al. 2013). The emergence and cancellation of magnetic flux may also cause the formation or disappearance of the barb magnetic structures (Li \& Zhang 2013). The larger-scale patchy counter-streaming in EUV along the filament channel from one polarity to the other observed by Hinode/EIS was found to be related to the intensity of the plage or active network (Chen et al. 2014).

Up to now, there are two viewpoints on the counter-streaming flows:
One is the steady bidirectional streaming everywhere in the filament along adjacent closely spaced threads (Zirker et al. 1998);
The other is the larger-scale patchy counter-streamings in EUV along the filament channel from one polarity to the other (Chen et al. 2014).
Chen et al. (2014) have not found EUV counterpart of the fine-structured H$\alpha$ counter-streaming along the filament channel implied by Zirker et al. (1998). They explained that the ubiquitous H$\alpha$ counter-streamings found by previous researchers are mainly due to longitudinal oscillations of filament threads, which are not in phase with each other. The larger-scale patchy counter-streaming flows are another component of unidirectional flows inside each filament thread in addition to the longitudinal oscillation. Therefore, the nature of the material flows in the filament threads deserves further investigation.
In addition, although the structures of quiescent filaments have also been studied for decades,
the fine-scale structures of filament barbs and spines are still essentially unclear at present.
To address these issues, we present two quiescent filaments observed by New Vacuum Solar Telescope to show small-scale structures
and their evolution of filament barbs and the material flows in the filament threads.

\begin{figure}
\centering
\includegraphics[width=13cm, angle=0]{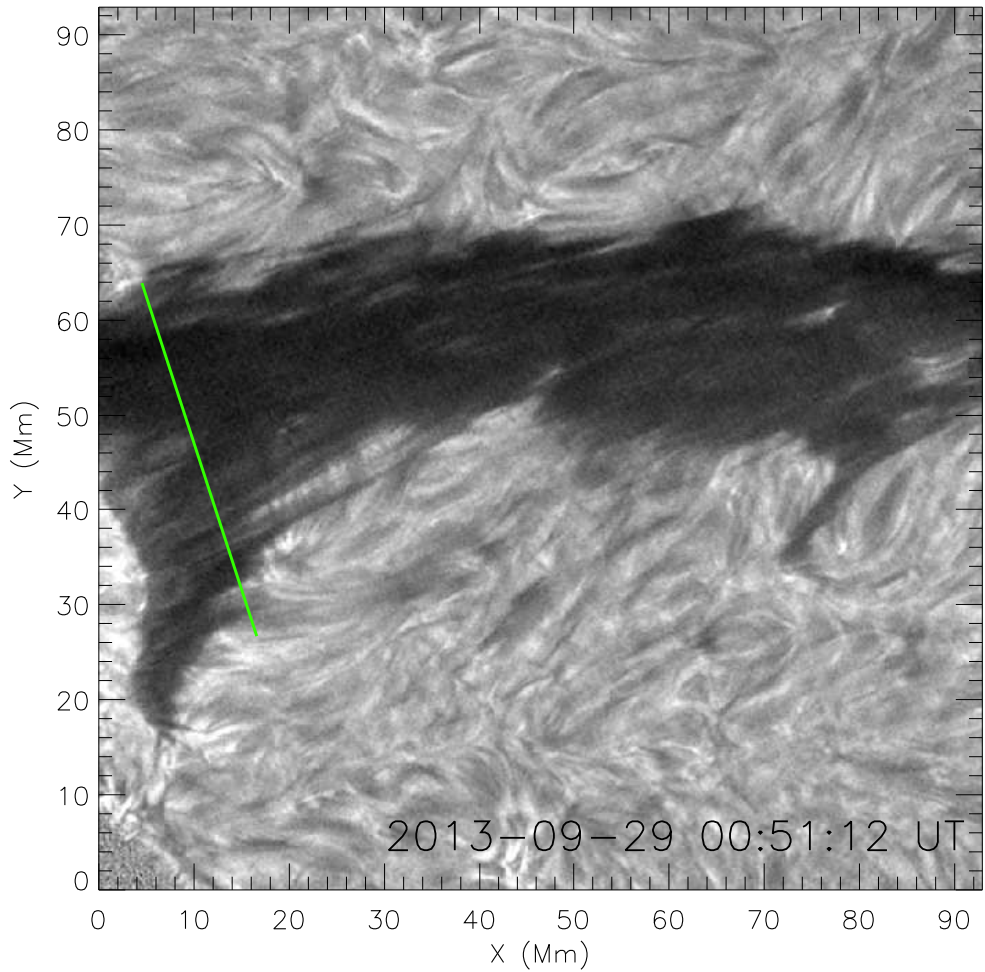}\\
\caption{High-resolution H$\alpha$ image of the quiescent filament observed by NVST at 00:51:12 UT on 2013 September 29. 
The green line is used to mark the position of the time slice of Fig. 3.}
\label{Fig1}
\end{figure}

\section{Observation and data processing}

The New Vacuum Solar Telescope (NVST) is a vacuum solar telescope with 985 mm clear aperture. The vacuum system consists of two vacuum tubes as the telescope should rotate on its altitude axis in addition to its azimuth axis. These two vacuum tubes are separated by two vacuum windows. An optical window (W1) with a diameter of 1.2 meters is placed on the top of the vacuum tube to keep the air pressure inside the tube lower than 70 Pa. The optical system after W1 is a modified Gregorian system with an effective focal length of 45 m. The primary mirror is a parabolic imaging mirror with a clear aperture of 985 mm. The detailed information about NVST can be seen from the paper of Liu et al. (2014). It is the primary observational facility of the Fuxian Solar Observatory (FSO). The main scientific goal of NVST is to observe the fine-scale structures in both the photosphere and the chromosphere. A multi-channel high resolution imaging system was set up and one chromospheric channel and two photospheric channels can be used to observe now. The tracking accuracy of NVST is less than 0.3 arcseconds. The H$\alpha$ filter is a tunable Lyot filter with the bandwidth of 0.25 \AA. Thanks to the good atmospheric seeing of Fuxian lake, the observations at NVST can last for several hours. The data used in this study were obtained by the NVST in H$\alpha$ 6562.8 \AA\ from 00:51:12 UT to 04:12:08 UT on 2013 September 29 and from 05:53:07 UT to 08:07:56 UT on 2012 November 2. The cadence of H$\alpha$ observations is 12 s and the pixel size is 0.16$^\prime$$^\prime$. The reduced data from NVST are classified into two levels. One is the Level 1 data that are processed by frame selection (lucky imaging) (Robert 2003). The other is the Level 1+ data that are reconstructed by speckle masking (Weigelt 1977; Lohmann et al. 1983) or iterative shift \& add (ISA; Liu et al. 1998). The H$\alpha$ data used in this paper are dark current subtracted and flat field was corrected to Level 1, and then are reconstructed to Level 1+ with the speckle masking method of Weigelt (1977). The images were co-aligned by using the subpixel registration algorithm (Feng et al. 2012; Yang et al. 2014).

\begin{figure}
\centering
\includegraphics[width=13cm, angle=0]{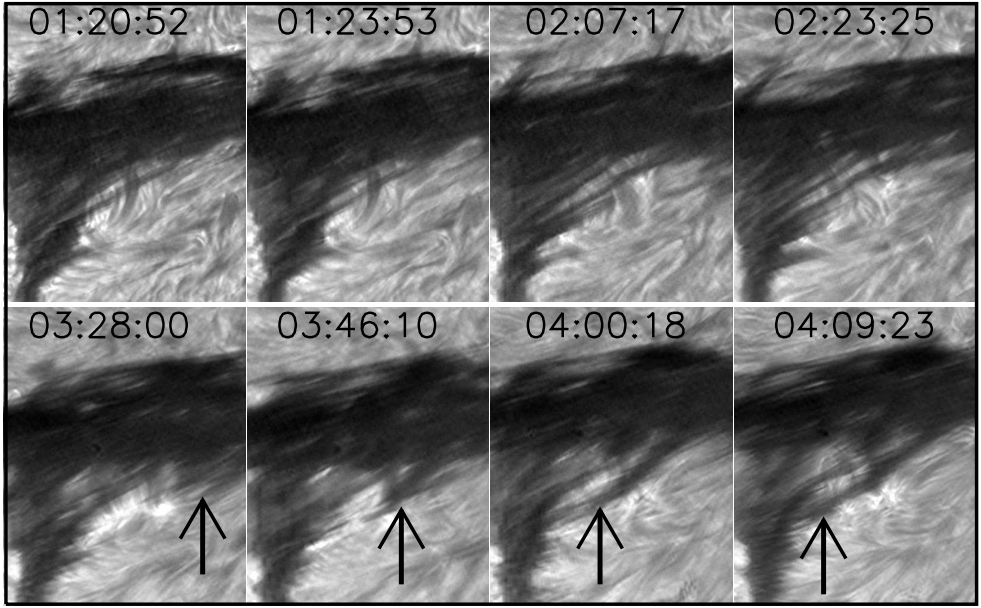}\\
\caption{Sequence of high-resolution H$\alpha$ images to show the evolution of the filament barb from 01:20:52 UT to 04:09:23 UT on 3013 September 29. 
The black arrows indicate the materials, which are flowing from the spine to the barb.}
\label{Fig2}
\end{figure}

\begin{figure}
\centering
\includegraphics[width=13cm, angle=0]{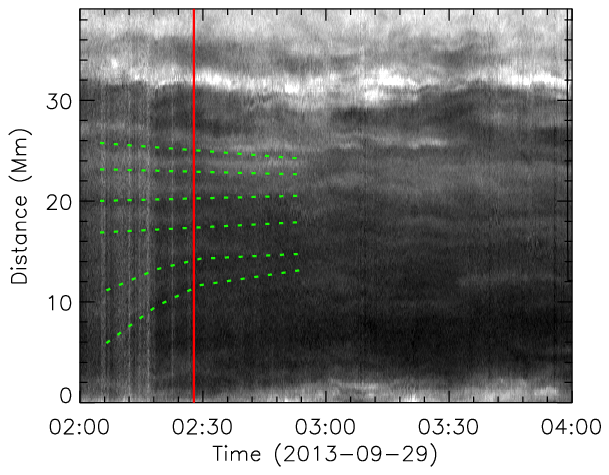}\\
\includegraphics[width=12cm, angle=0]{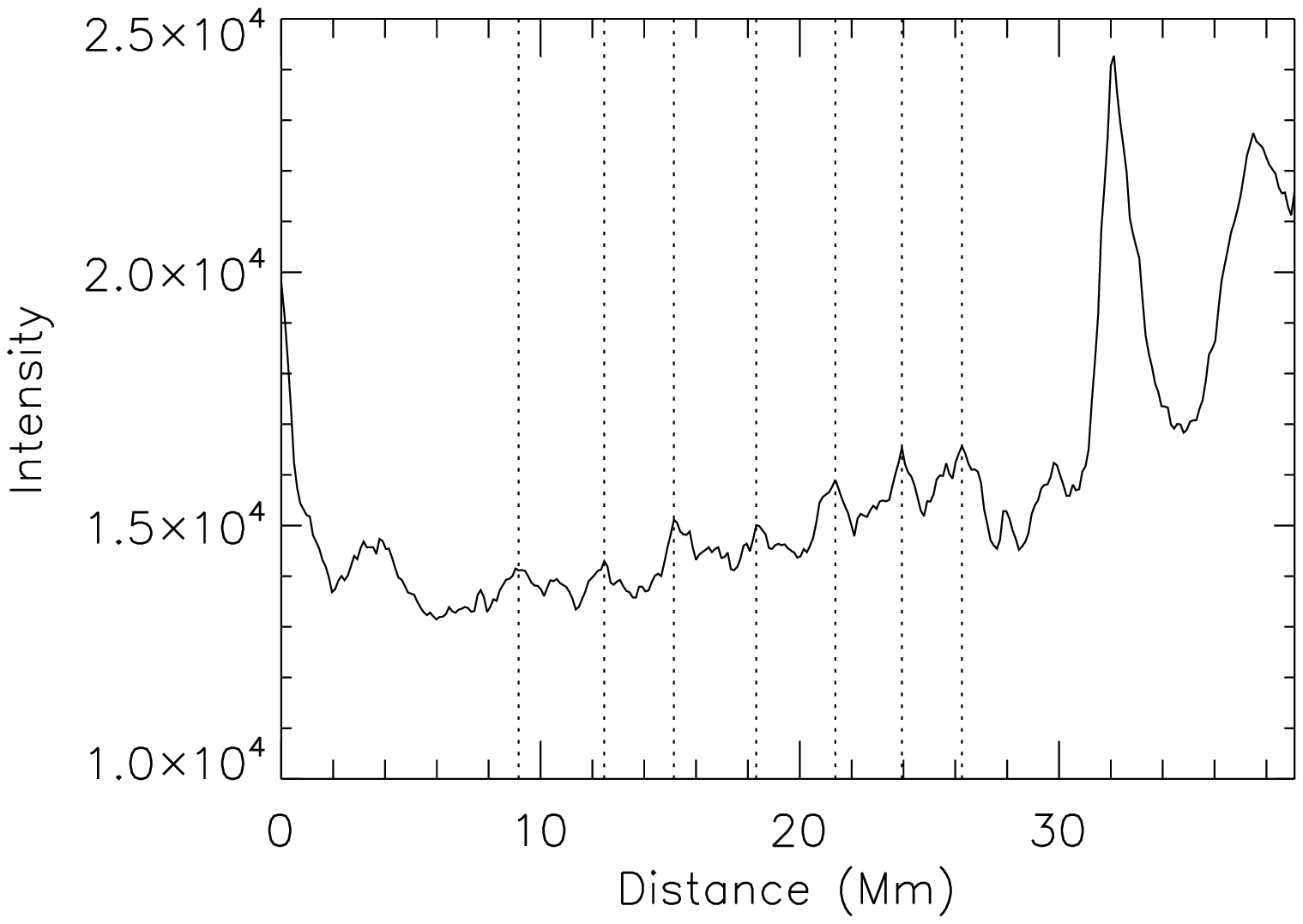}
\caption{Upper panel: A time-slice taken from high-resolution H$\alpha$ images at the position marked by the green line in Fig. 1. 
The green dotted lines indicate the six parallel flux tube-shaped structures formed in the barb. 
Lower panel: The intensity along the red line marked by the red line in the upper panel.
The vertical dotted lines outline widths of the six tube-shaped structures were formed during the barb evolution.}
\label{Fig3}
\end{figure}

\begin{figure}
\centering
\includegraphics[width=13cm, angle=0]{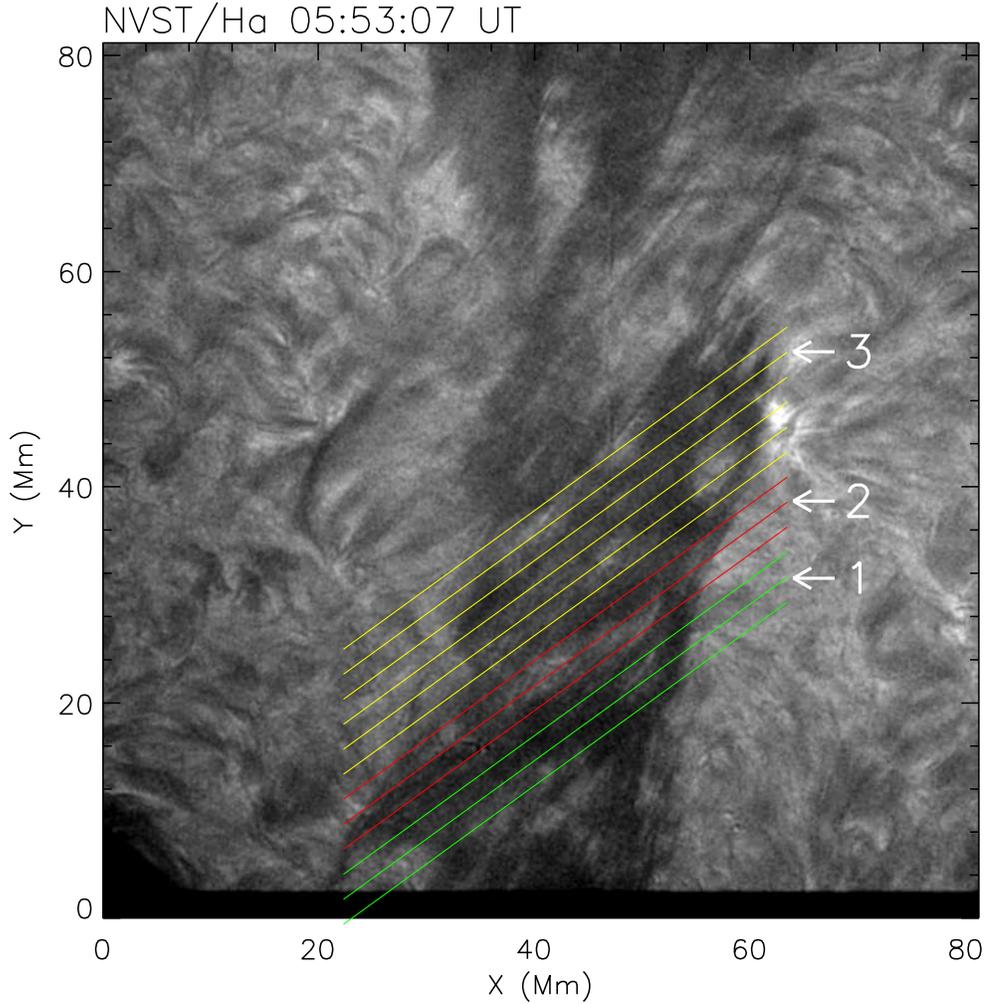}
\caption{High-resolution H$\alpha$ image of the quiescent filament observed by NVST at 05:53:07 UT on 2012 November 2. 
The green and yellow lines marked the areas, which have material flows with the same direction and the red lines denote the opposite material flows. 
The lines marked by the numbers indicate the positions of the time slice of Fig. 5. }
\label{Fig4}
\end{figure}

\begin{figure}
\centering
\includegraphics[width=13cm, angle=0]{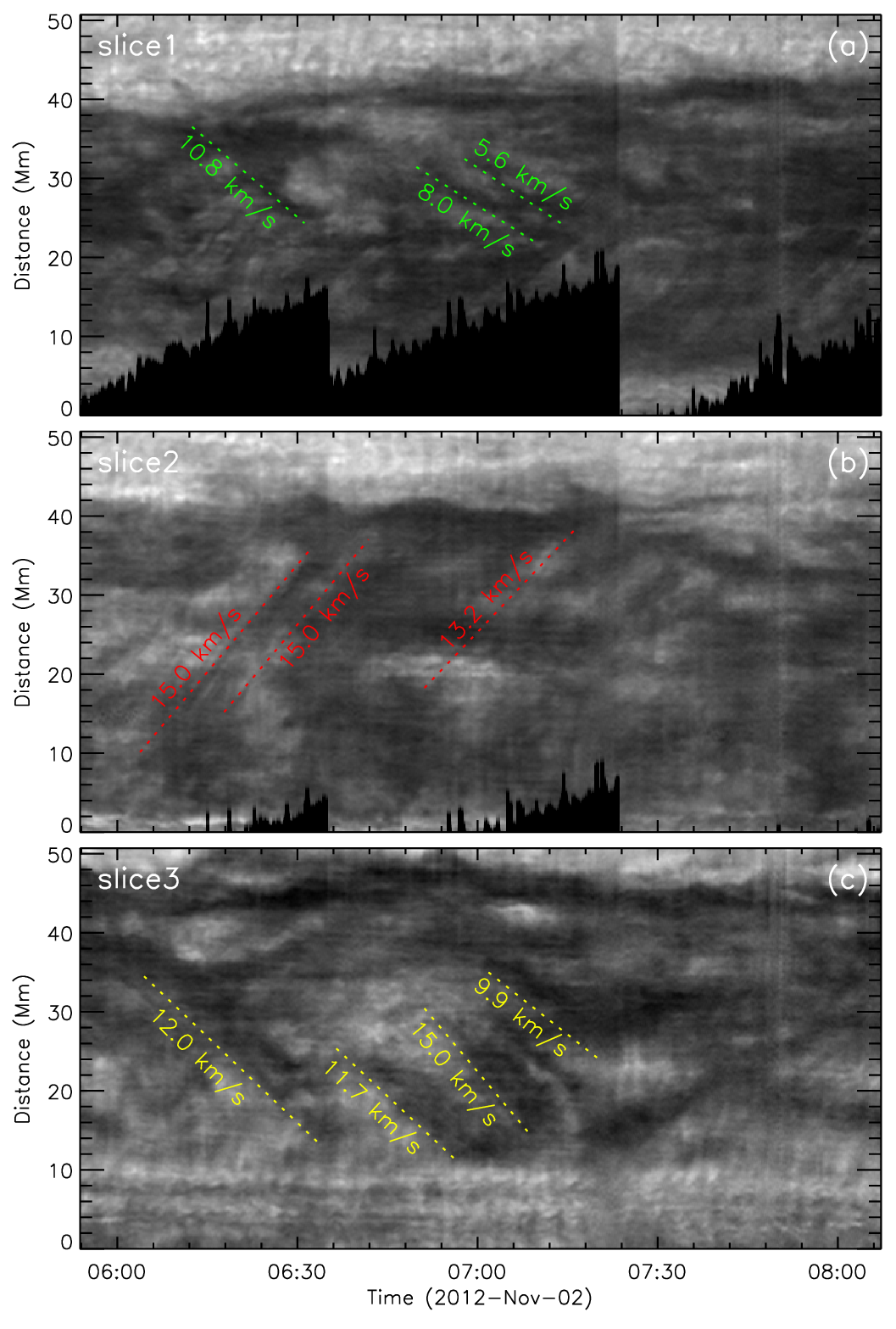}
\caption{The three time slices taken from high-resolution H$\alpha$ images at positions marked by the numbered lines in Fig. 4. 
The velocities of the counter-streaming are marked in the figure.}
\label{Fig6}
\end{figure}

\section{Result}

Figure 1 shows the high-resolution H$\alpha$ images of the quiescent filament observed by NVST at 00:51:12 UT on 2013 September 29. The quiescent filament was located at the northwest of the Sun. The centre of the field of view was located at x=240$^\prime$$^\prime$ and y=450$^\prime$$^\prime$. Due to the field of view, only a part of the quiescent filament was observed. The quiescent filament was a very mature large one with a large barb and spine. This filament appeared at the east of the Sun since 2013 September 20 by checking the H$\alpha$ data of BBSO. The threads of the barb smoothly transited into the spine. Moreover, the boundary between the barb and the outer atmosphere is very neat. To address the special structures of the barb, we measure the intensity of the barb and the outside atmosphere. The difference of the intensity between both sides of the boundary of the barb is about 10\%. It illustrates that each side of the boundary has different magnetic structures. Up to now, we have not found the explanation about why the boundary between the barb and the outer atmosphere is so obvious. We conjecture that this structure may be related to the special magnetic structures of the barb. When quiescent filaments appear at the solar limb, the bright arched structures are often observed (see Fig. 3b and 3c in the paper of Mackay et al. 2010). Many vertical threads are connecting the boundary of the arched structures and the spine of the prominences. Moreover, there is a cavity below the arched magnetic structure. We suspect that the obvious boundary between the barb and the outside may be due to the projection of the arched structures of the quiescent filaments on the Sun. This speculation need continuous high resolution observations to trace the evolution of solar filaments from the solar disk to the limb.

Figure 2 shows the evolution of the barb from 01:20:52 UT to 04:09:23 UT. At the beginning, the barb is composed of parallel narrow threads, similar to those in the spine. The individual threads are gradually blending with neighboring threads. After half an hour, the intensive narrow threads evolved into the distinct parallel tube-shaped structures.  The structures maintained for about half an hour and began to merge together, accompanied with the material flow from the spine to the barb. The black arrows in Fig. 2 indicate the material flows from the spine to the barb. The typical width of the threads in the quiescent filaments is between 0.2 and 0.3 arcseonds (Lin et al. 2005). The pixel size of H$\alpha$ in the Level 1+ data is 0.16$^\prime$$^\prime$. The resolution of NVST can identify the threads of the quiescent filament. The width of the parallel tube-shaped structures was larger than the typical size of the individual threads. This evolution process of the filament can be seen from the supplementary movie 1.

In order to show the change of the barb, we make a time slice along the green line in Fig. 1. The time slice is shown in the upper panel of Fig. 3. The length of the slice is 40 Mm and the time is from 02:00 UT to 04:12 UT. The green dotted lines indicate the six parallel flux tube-shaped structures formed in the barb. The six parallel tube-shaped structures had a convergence motion and merged each other gradually. The change of the intensity along the red line in the upper panel of Fig. 3 can be seen from the lower panel of Fig. 3. The vertical dotted lines are marked to show the width of the six parallel tube-shaped structures. The widths of the structures are 3.3, 2.7, 3.2, 3.1, 2.6, and 2.3 Mm. The average width is about 2.9 Mm. The flowing plasma along the filament threads may be taken as evidence for alignment of threads with the local magnetic field. This scenario is suitable for the flows in the horizontal magnetic fields of the quiescent filament. However, for the flows in the vertical structures of the quiescent filaments, the flows may be not field-aligned. Because the velocities of the flows in the vertical threads of the quiescent filaments are lower than free-fall velocities. Chae et al. (2008) explained the vertical threads of the quiescent prominence in terms of magnetic dips in the initially horizontal magnetic fields. To our knowledge, the formation of the regular arrangement tube-shaped structures during the evolution of the barb have not been reported. The role of the tube-shaped structures in the formation and the evolution of the barb is unclear. Please keep in mind that we just observed these structures and the exact direction of magnetic fields may not align with these structures for the lack of vector magnetic field data. Even though the weather is not very well on 2013 September 29, the evolution of the barb can be seen clearly. However, the material flows in the filament threads of the filament have not been observed.

To study the material flow in the filament, another quiescent filament on 2012 November 2 is presented. This quiescent filament is located close to the south pole. The centre of the field of view was located at x=-300$^\prime$$^\prime$ and y=-743$^\prime$$^\prime$. Figure 4 shows the high-resolution H$\alpha$ image of the quiescent filament at 05:53:07 UT observed by NVST. The filament is composed of the threads in the spine. The angle between the spine and the threads is about 36 degrees. From the evolution of the filament, we found that the material flow in the filament threads is different from the results obtained by some previous researchers (Zirker et al. 1998; Lin et al. 2003). Zirker et al. (1998) and Lin et al. (2003) found that the steady bidirectional streaming everywhere in the filament along adjacent closely spaced threads. However, we found that the counter-streaming flows in this filament did not obey the rule that the flows in the adjacent threads have opposite directions. The flows in one patch of the filament threads have the same direction and the flows in the adjacent patch of the filament threads have opposite directions. This finding is consistent with the large-scale counter-streaming flows in EUV filament channel observed by Chen et al. (2014).

To calculate the exact velocity of the material flows, we make one time slice every 5 pixels along the filament spine, which is parallel to the threads of the filament with an inclination angle 36 degrees. In Figure 4, we selectively draw 12 different color lines every 20 pixels to show the different regions with material flows in opposite directions. Along the filament spine, we make forty time slices between the uppermost yellow line and the lowermost green line in total to calculate velocities of the material flows. Lin et al. (2009) found that the individual threads swayed back and forth sideways in the plane of the sky at quite small amplitude ($<$ 100 km). In order to eliminate the swaying motion of the threads, we take the average intensity of the three pixels perpendicular to the line, in which one pixel is on the line and the other two pixels are on both sides of the point. We found that there was a large-scale counter-streaming with a certain range in the filament. The same color lines indicate that the material flows along these lines have the same direction. We give three examples to show the material flows with time. The lines 1, 2, and 3 are parallel to the threads of the filament and overly the threads. We make three time slices along this three positions to show the mass flows with time. Time slice1, slice2, and slice3 along the three lines marked by 1, 2, and 3 in Fig. 4 are shown in Fig. 5. We trace the features to estimate the velocity along the threads by using linear fitting. The velocities of counter-streamings range from 5.6 km/s to 15.0 km/s. This evolution process of the filament can be seen from the supplementary movie 2.

\section{Conclusion and Discussion}

The small-scale structures of filament barbs and counter-streaming in the threads are investigated in detail by using high resolution H$\alpha$ data. First, we found that the thin threads in the barb can form the parallel tube-shaped structures with the width about 3 Mm. These structures formed during the evolution of the barb. Following the material flows from the spine to the barb, these structures merged and disappeared; Second, we found the large-scale counter-streaming with a certain range along the threads in the filament spine rather than the steady bidirectional counter-streamings everywhere in the filament along adjacent closely spaced threads. The velocities of the counter-streaming range from 5.6 km/s to 15.0 km/s.

The thin dark threads are along the quiescent filament spines and barbs observed by Swedish 1-m Solar Telescope (SST) and Dutch Open Telescope (DOT) (Heinzel 2007). These threads are taken as the shear of the magnetic field lines. The flowing plasmas along the thin threads are taken as evidence from alignment of threads with the local magnetic field (Martin et al. 2008). Chae et al. (2005) suggested that filament barbs are cool matter suspended in local dips of the magnetic field lines formed by magnetic reconnection in the chromosphere. However, the detail evolution of the barbs has not been investigated by previous authors. The barb investigated in this paper is very typically mature one. The boundary between the barb and the surrounding atmosphere is very neat. The threads of the barb smoothly connected the spine of the filament. Unlike the filaments studied by Lin et al. (2005, 2007), the parallel threads of the filament studied in this paper have an angle of about 36 degrees with the spine of the filament. The parallel tube-shaped structures formed during the barb evolution, which has not been reported in the previous paper. The same phenomenon did not occur in the spine of the filament during its evolution. The width of these structures in the barb ranges from about 2.3 Mm to 3.3 Mm, which is larger than the width of the threads (150-450 km) obtained by Lin et al. (2005). The nature of these parallel structures deserves further investigation. Because the velocities of the flows in the vertical threads of the quiescent filaments are lower than free-fall velocities, Chae et al. (2008) and Chae (2010) proposed magnetic dips in initially horizontal magnetic fields to explain this phenomenon. We just observed downflow from the spine to the barb. As this filament was observed in the solar disk, the upflows in the vertical structures of the prominences found by Berger et al. (2010) can not be observed. Joshi et al. (2013) found that the formation and disappearance of the filament barbs is closely related to the material flows. Our results also confirm this viewpoint. However, why the parallel tube-shaped structures appear during the evolution of the barb and the role in the barb formation and evolution need more observation and simulation to study.

The steady bidirectional streaming everywhere in the filament along adjacent closely spaced threads was discovered by Zirker et al. (1998). They found that the counter-streaming is especially clear in the spine, but the same phenomenon also exists in the barbs. Therefore, the counter-streamings are also evidenced by several authors to exist in the filament (Lin et al. 2003; Schmieder et al. 2008). Ahn et al. (2010) found that the pattern of horizontal counter-streaming motions is due to the return of moving plasma fragments. However, from our observation, we have not found the ubiquitous H$\alpha$ counter-streamings in adjacent threads of the filament reported by previous researches (Zirker et al. 1998; Lin et al. 2003; Schmieder et al. 2008). We found that the large-scale counter-streamings existed in the filament spine. One patch of the flows has the same direction and the adjacent patch of the flows has opposite directions. The width of the patches is about ten arcseconds.  The patches exhibited alternately along the spine of the filament. This result is different from that of Zirker et al. (1998). Chen et al. (2014) used the extreme ultraviolet (EUV) spectral observations and found there are no EUV counterparts of the H$\alpha$ counter-streamings in the filament channel. The larger-scale patchy counter-streamings in EUV along the filament channel were found from one polarity to the other (Chen et al. 2014). The upflows and downflows observed by Chen et al. (2014) were about 10 km/s, which are the same as the transverse velocities along the filament threads obtained by us. They suggested that the ubiquitous H$\alpha$ counter-streamings found by previous researchers are due to the longitudinal oscillations of the filament threads, which are not in phase with each other. The results mentioned above are obtained by using the line-of-sight velocity or Doppler velocity and have not traced the transverse motion of the features of the filament observed by H$\alpha$ observation.

From our observation, we found that the dark material flows are well co-aligned with the threads and do not cross the threads. Moreover, our results show that there is a large-scale counter-streaming with a certain range in the threads of the quiescent filament. Ubiquitous H$\alpha$ counter-streamings in adjacent threads of the filament are not found in our example. The long oscillation period of a quiescent filament is 98 minutes obtained by Ning et al. (2009a). The continuous material flows along the threads lasted for about two hours and fourteen minutes and did not change their directions during our observation. If the material flows are caused by the transverse oscillation of the filament, we should observed the returning flows. Lin et al. (2009) and Lin et al. (2011) found that the period of the swaying motions of individual filament threads is about 5 minutes by using high-resolution observations obtained by the Swedish 1m Solar Telescope in La Palma. The same period was also obtained by Ning et al. (2009b) by analyzing small-scale oscillations in a quiescent prominence observed by Hinode. Parenti (2014) presented a detailed review for the oscillation of the prominences.  Moreover, the amplitude of the transverse oscillation is very small, about 5 Mm (Ning et al. 2009b). The continuous material flows can last for the whole observational time and did not change their direction. Therefore, we think that these flows are not caused by the transverse oscillation of the filament. Our result supports the viewpoint suggested by Chen et al. (2014) and Okamoto et al. (2007) that ubiquitous H$\alpha$ counter-streamings in adjacent threads of the filament may be driven by the longitudinal oscillation of the filament threads. Berger et al. (2010) found that the prominence plasma is entrained by the upflows, which have ascent speeds of 13 -17 km/s. The fine structures acquired at Ca II H have the rising speed of 5-20 km/s before the filament formation (Okamoto et al. 2010). Some of them fell down after reaching the maximum heights, while the others went up with decelerating speed and stayed in the corona. The upflows may counter with the ubiquitous downflow ones in the prominence (Berger et al. 2008). The detail relationship between the downflow/upflow and the transverse material flows in the threads of the quiescent filament deserves further investigation.

\begin{acknowledgements}
The authors are grateful to the referee for his/her constructive suggestions. This work is supported by the National Science Foundation of China (NSFC) under grant numbers 11373066, 11373065, 11203077, Yunnan Science Foundation of China under number 2013FB086, the Talent Project of Western Light of Chinese Academy of Sciences, 
the National Basic Research Program of China 973 under grant number G2011CB811400. Key Laboratory of Solar Activity of CAS under number KLSA201303, KLSA201412, KLSA201407. The authors thank Rui Wang, Dingchang Wang for their observation.
\end{acknowledgements}

\label{lastpage}

\end{document}